# Pump-Push-Probe for Ultrafast All-Optical Switching: The Case of a Nanographene Molecule

Giuseppe M. Paternò[1*], Luca Moretti[2], Alex J. Barker[1], Qiang Chen[3], Klaus Müllen[3], Akimitsu Narita[3], Giulio Cerullo[2], Francesco Scotognella[1,2] and Guglielmo Lanzani[1,2*]

[1]Istituto Italiano di Tecnologia, Center for Nano Science and Technology, Milano, 20133, Italy
[2]Politecnico di Milano, Department of Physics, Milano, 20133, Italy
[3]Max Planck Institute for Polymer Research, Mainz, 55128, Germany

## Abstract

In the last two decades, the three-beams pump-push-probe (PPP) technique has become a well-established tool for investigating the multidimensional configurational space of a molecule, as it permits to disclose precious information about the multiple and often complex deactivation pathways of the excited molecule. From the spectroscopic point of view, such a tool has revealed details about the efficiency of charge pairs generation and conformational relaxation in *p*-conjugated molecules and macromolecules. In addition, PPP has been effectively utilised for modulating the gain signal in conjugated materials by taking advantage of the spectral overlap between stimulated emission and charge absorption in those systems. However, the relatively low stability of conjugated polymers under intense photoexcitation has been a crucial limitation for their real employment in plastic optical fibres (POFs) and for signal control applications. Here, we highlight the role of PPP for achieving ultrafast all-optical switching in *p*-conjugated systems. Furthermore, we report new experimental data on optical switching of a newly synthesised graphene molecule, namely dibenzo[*hi,st*]ovalene (DBOV). The superior environmental and photostability of DBOV and, in general, of graphene nanostructures can represent a great advantage for their effective applications in POFs and information and communications technology.

[*] Correspondence should be addressed to Giuseppe M. Paternò giuseppe.paterno@iit.it, or to Guglielmo Lanzani Guglielmo.lanzani@iit.it.

# 1. Introduction

Back in 1990 the scientific community working on organic semiconductors was engaged in a large research effort aimed at identifying new non-linear optical materials for all-optical communication. The funding notion was the observation that large *p*-electron delocalization, typical in conjugated polymers and large molecules, should be associated to large susceptibility values. In turn non-linear terms of susceptibility, typically third order, could be implemented in a number of photonic devices expected to be efficient and fast. In a decade excellent results were obtained in fundamental knowledge, essentially dealing with the synthesis of new compounds, the understanding of structure-property relationship, and the realization of advanced spectroscopy experiments, yet practical achievements suitable for further development into real applications did not arrive. Organic semiconductors in different shapes indeed demonstrated very large non-linear optical coefficients, but their stability under optical pumping, optical guiding properties (e.g. optical losses) and spectral range of operation (often limited to the visible) were not appropriate for information and communications technology (ICT). Slowly, the interest toward this thematic tuned down while electroluminescence for organic light-emitting diodes[1] and later photovoltaics[2] emerged as new frontiers for material development. More recently, ultrafast all-optical switching was reported in solid state amplifiers, plastic optical fibres, optofluidic channels and lasers based on resonant excited state effects[3]. Essentially, the detrimental absorption associated to charge carriers that spectrally overlaps with stimulated emission hampering electrical pumped lasing in most organic semiconductors, was exploited for modulating optical gain[4]. Because charge carrier lifetime can be modulated by controlling intermolecular interactions, such induced quenching could be made extremely fast, being a transient switch of the gain signal. Such all-optical control is fast (sub-picosecond) and large (easily modulating 100% of the gain signal), in spite of being resonant. This breaks the common knowledge fast-small-nonresonant *vs*. slow-large-resonant. However, it does not circumvent the problem of energy management, as for in a resonant process a lot of energy is deposited in the materials and requires efficient dissipation. As a matter of fact, this limit the modulation rate several orders of magnitude below the theoretical limit. In this context, new 2D materials, in particular graphene molecules[5], appear as interesting candidates for readdressing the challenge, as those materials have very fast excited state dynamics and usually rather high environmental stability.

In this feature article, we provide a summary of recent advances on pump-push-probe technique and its application to ultrafast all-optical switching in *p*-conjugated systems. In addition, we report new interesting results on ultrafast switching of stimulated emission signal in a newly synthesised stable graphene molecule, which holds great promise for possible future applications of these novel advanced functional materials in photonics and ICT.

## 2. The pump-push-probe experiment

In terms of wave interaction picture, the pump-push-probe (PPP) experiment is a nonlinear fifth order process. However, in the well separated pulses regime this experiment is simply accounted for by a double modulation picture. First, the pump pulse places population into the excited state. Second, the push pulse displaces part of this population to a higher lying excited state. Then internal conversion can either simply bring back the population to the initial excited state or it can take a different path, reaching another excited state. In this case deactivation back to the initial excited state might become longer then direct internal conversion. In a system with inversion symmetry, a one photon transition with energy equal to the sum of pump and push photons reaches an excited state that is different from that reached by the push after the pump (two step transition). This difference comes from the dipole selection rule, that couples opposite symmetry states. Therefore, deactivation can occur along very different paths in the two situations because it involves states of different symmetries, albeit almost equienergetic. This demonstrates that the PPP technique allows exploring the multidimensional configurational space of a molecule. In particular, it can identify branching routes in the deactivation path of excited states. Often such branching points lead to charged states that requires a minimum excess energy to be accessible. The opening of different decay paths above a certain energy cause the breakdown of the Vavilov rule. In conjugated polymers like poly(p-phenylene vinylene) (PPV)[6] or ladder-type poly(para)phenyl (m-LPPP)[7] and poly(9,9-dioctylfluorene) (PFO)[8] the PPP technique has been used to identify autoionizing states (figure 1a,b). In fluorene oligomers the PPP technique allowed to observe ultrafast non adiabatic planarization following the push re-excitation (figure 1c)[9]. In photonics the PPP technique is the natural tool for investigating optical switching, for the push pulse act as a modulation on the probe transmission[10]. An alternative to a push induced electronic transition is the push induced vibrational transition, when the push photon energy is in the vibrational range, typically mid-IR[11]. In this configuration the push effect is like a local heating, bringing excess vibrational energy to the initial excited state.

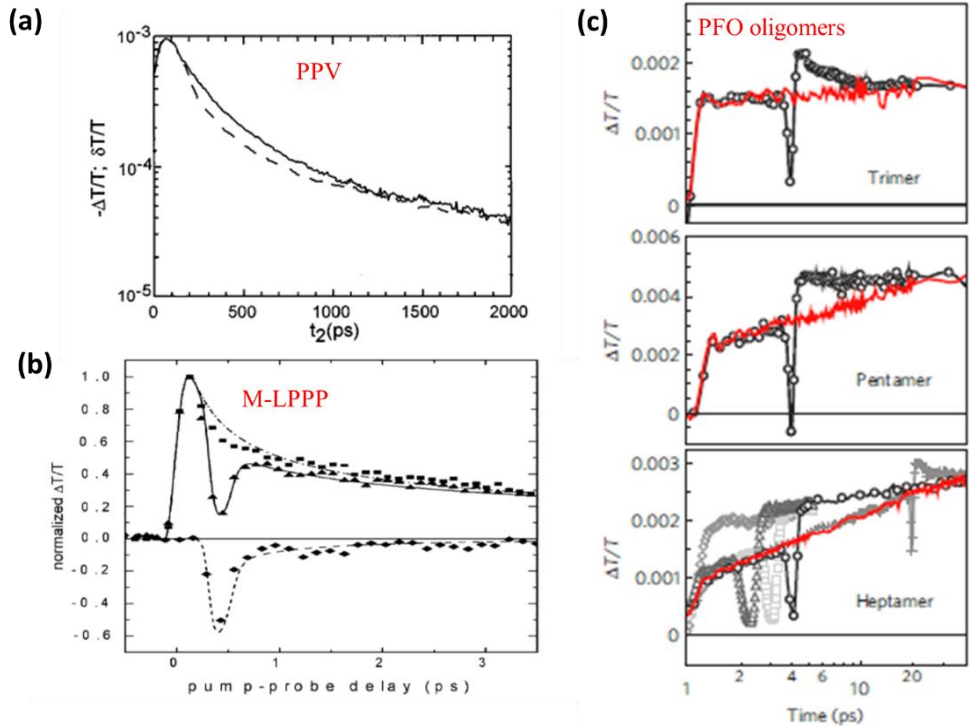

**Figure 1.** (a) Pump-probe (solid line) and pump-push-probe (dashed line) dynamics for PPV in the gain region. This study (Frolov et al.[6]) reported for the first time the role of autoionizing states in conjugated polymer, and the possibility to exploit those states to modulate the stimulated emission signal. (b) Pump-probe (■) and pump-push-probe (▲) dynamics for the ladder polymer m-LPPP. Here, Gadermaier et al.[7] investigated the charge generation efficiency and the fate of such photogenerated species by means of pump-push-probe technique. (c) Dynamics of the stimulated emission region for PFO oligomers with (○) and without (red line) push re-excitation. In this study, Clark et al.[9] attributed the "overshoot" in the stimulated emission dynamics after push re-excitation to ultrafast torsional relaxation. Figure 1a is re-adapted from ref.[6] and figure 1b from ref[7]. under permission of the American Physical Society (APS). Figure 1c is re-adapted from ref.[9] under permission of Springer Nature.

## 3. Pump-push-probe for all-optical switching

A number of experimental studies have shown the role of charge carriers as intrinsic loss in the optical gain process of conjugated materials[12]. Such a competition arises from the fact that charged states have optical transitions just below the neutral absorption edge (figure 2), and strongly overlapping with stimulated emission (SE). SE gets eventually overtaken by photoinduced absorption (PA) in the solid state where charge generation due to intermolecular effects become significant (figure 3a). Therefore, this phenomenon hampers the occurrence of electrically pumped lasing action for organic conjugated systems, as in this case charge absorption is massive and unavoidable and overwhelms optical gain dramatically.

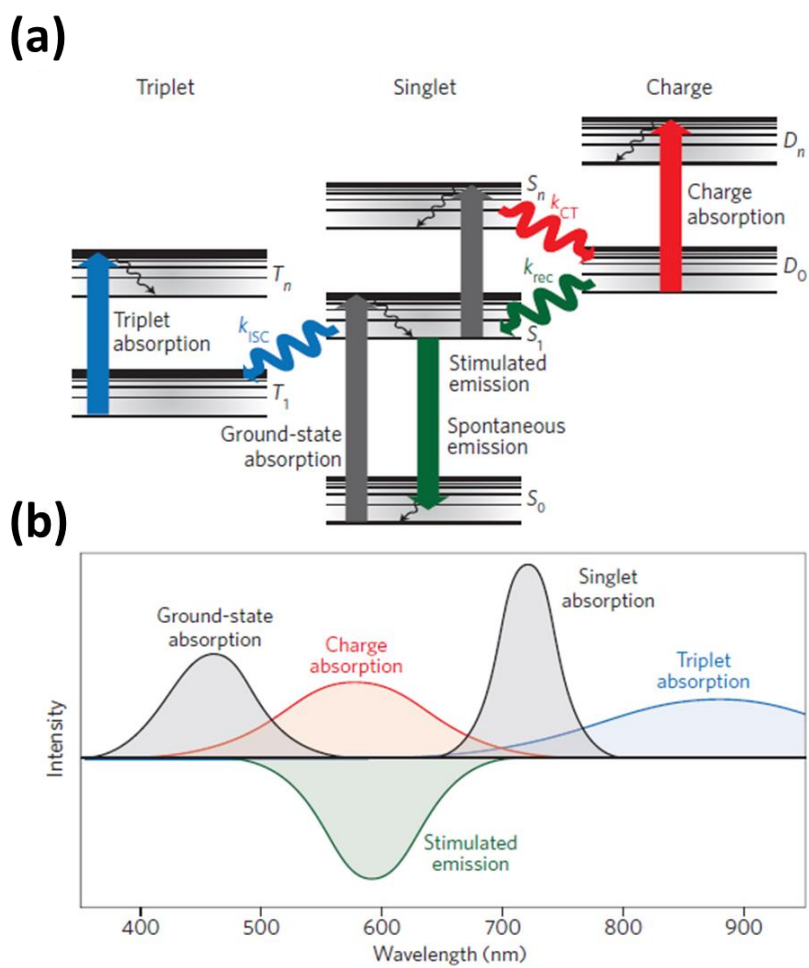

**Figure 2.** (a) Energy level architecture of organic semiconductors with single, triplet and charge excitons. (b) Absorption bands related to the abovementioned transitions. It is worth noting the spectral overlap between stimulated emission and charge absorption. This figure has been taken from ref. [4] under permission of Springer Nature.

On the other hand, selective excitation of the charged states can be effectively utilized for optically switching on/off the SE signal. However, to achieve charges recombination and hence recovery of SE in the picosecond time-regime, it is necessary to exploit the intrinsic one-dimensional character of conjugated chains as isolated and electronically confined elements, in which photogeneration of long-lived interchain charge pairs is suppressed. For instance, previous studies on isolated conjugated polymer chains have shown evidence of early events (< 300 fs) of intrachain charges excitation[13]. In this case, charge pair decays geminately in the ultrafast time regime (≈ 500 fs) to the lowest singlet state, due to the strong confinement of such excitations within one single polymer chain. This interesting property of isolated conjugated polymer chains can be thus exploited for the development of an ultrafast photonic switch based on a three-beams geometry[14]. Here, the pump pulse populates $S_1$ while the push pulse at some later time depletes $S_1$, thus reducing the optical gain experienced by the probe pulse. In addition, as anticipated in the previous paragraph, the push induced re-excitation to $S_n$ promotes excitons dissociation into charge pairs[7, 15]. This eventually results in an increase of the PA associated to intramolecular charge pairs and consequent quenching of the SE signal, as reported in figure 3b[14]. Interestingly, recovery occurs within 1 ps, a time-scale that is considerably larger than internal conversion ($\approx$ 50 fs)[16] and can be connected to intrachain charge recombination[17]. Therefore, the exploitation of PA generated by the resonant push pulse enhances the modulation depth in the SE signal at the cost of a longer recovery time, yet the process is fast enough to allow sub-ps responses. This in-fact demonstrates the possibility of achieving resonant ultrafast optical switch in conjugated polymers, with potential frequency in the THz regime. Furthermore, the switching process exhibits a relatively high on/off ratio that is typical for resonant-processes, coupled with an ultrafast response that conversely is unusual for such processes. However, some stability problems may arise due to the large amount of energy absorbed and quickly dissipated when the radiation is in resonance, thus limiting the actual operational switching rate. This critical issue justifies the search for new stable *p*-conjugated materials capable to undergo ultrafast resonant optical switching, which ideally show a higher resilience towards environmental stimuli and photoexcitation than conjugated polymers[18].

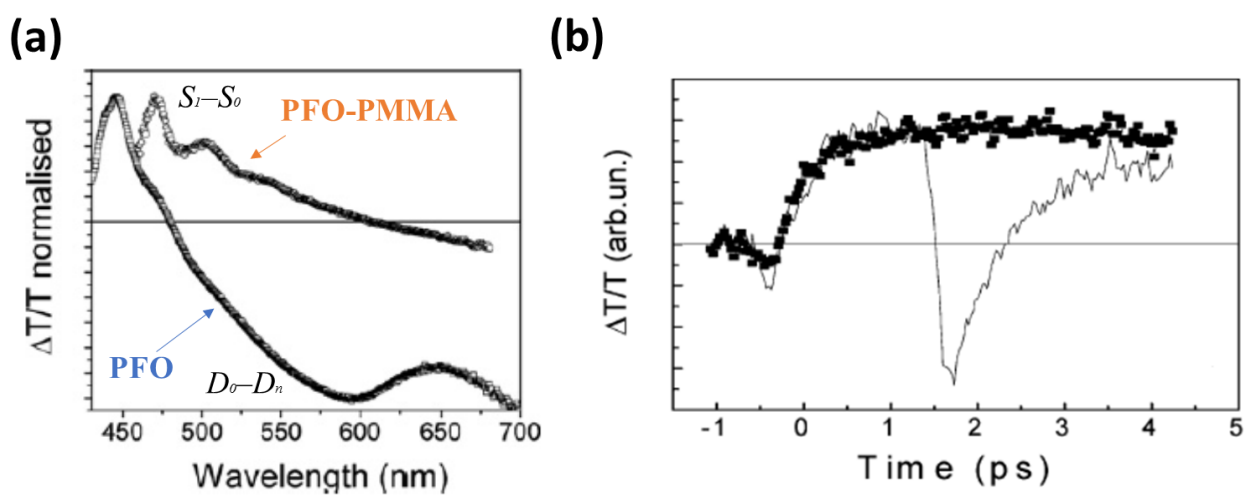

**Figure 3**. (a) Pump-probe spectra for pure PFO film and in blend with poly(methyl methacrylate) (PMMA). The broad negative band ranging from $\approx$ 490 nm to 650 nm corresponds to probe absorption due to intermolecular charge pairs ($D_0$-$D_n$). Note that such strong feature overwhelms stimulated emission ($S_1$-$S_0$) in the solid state, whereas the stimulated emission signal recovers when interchain effects are suppressed by dispersing PFO in polymer matrix. (b) Dynamics at a probe wavelength of 610 nm with push (−) and without push (■) for PFO dispersed in PMMA matrix. The pump-push delay is set at 1.5 ps. All the graphic material reported in this figure is adapted from ref.[14] under permission of the Royal Society of Chemistry (RSC).

## 4. Ultrafast resonant switching in a graphene molecule

The geometric confinement of graphene down to the molecular level into 1D (graphene nanoribbons, GNRs)[5, 19] and 0D (graphene quantum dots, GQDs)[5, 20] structures by means of bottom-up chemical synthesis has permitted to open a finite band-gap in its electronic structure, thus making possible the use of such structurally defined graphene fragments in the fields of optoelectronics and photonics[21]. Within the context of photonic applications, these extended π-conjugated systems have attracted growing interest very recently, due to their relatively high environmental stability and optical properties that can be easily tuned by modifying both size and edge configurations[22]. These are ideal properties for possible employment of graphene molecules in laser and signal control applications.

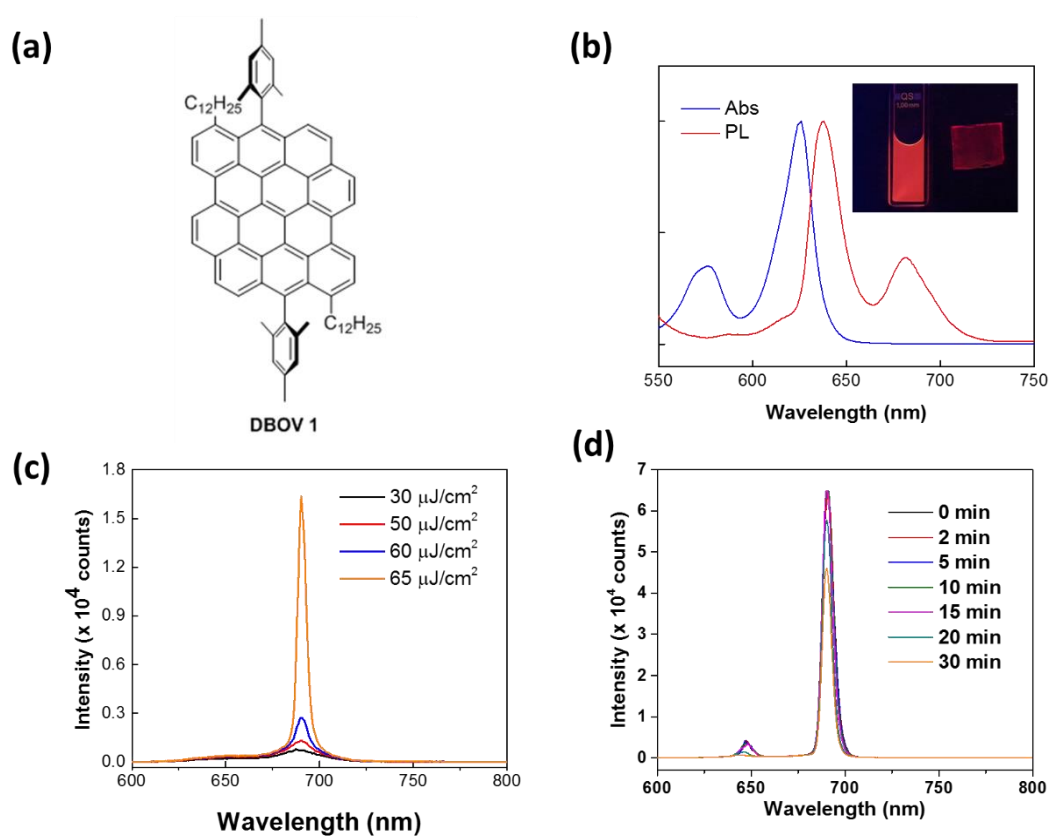

**Figure 4.** (a) Molecular structure of DBOV. (b) Normalised visible absorption and photoluminescence of DBOV in toluene solution (0.05 mg/mL). (c) ASE action of the DBOV:PS blend at 1 wt% (d) Time evolution of the ASE signal under irradiation with an excitation fluence five times higher the ASE threshold 320 μJ cm$^{-2}$) in air without any encapsulation. All the graphic material reported here have been adapted from reference[23] under permission of John Wiley & Sons.

We have recently characterised a novel graphene molecule, namely dibenzo[*hi,st*]ovalene (DBOV) (figure 4), with low-band gap, high photoluminescence quantum yield (79 % in solution) and small Stokes shift (< 10 nm) [23-24]. Furthermore, DBOV shows stable stimulated emission (SE) and amplified spontaneous emission (ASE, figure 4c) with remarkable environmental and operational stability, yielding only a 30% decay after 30 min of intense excitation in air with a pump fluence five times higher than ASE threshold (320 µJ cm$^{-2}$, 2×10$^6$ pulses, figure 4d)[23]. The pump-probe spectrum of DBOV in solution (figure 5a, top-graph) displays four main features, namely: i) a negative signal centred at 450 nm that can be attributed to PA from $S_1$ to $S_n$; ii) two positive peaks at 625 nm and 570 nm that can be connected to depletion of the ground state and its vibronic replica due to pump excitation PB; iii) a positive signal at 695 nm that can be ascribed to SE. However, if we pass from solution to solid film the SE signal at 695 nm is overwhelmed dramatically by a negative PA feature in the near infrared region within 200 fs (figure 5, central graph). On the other hand, dilution of DBOV in a polystyrene (PS) matrix (1 *wt*%) permittes to recover the SE and suppress such PA signal that, therefore, has to relate to the intermolecular distance experienced by the nanographenes (figure 5a, bottom graph). Interestingly, long-delay pump-probe experiments carried out on DBOV solution (figure 5b,c for spectra and dynamics) highlight the delayed kicking in of the PA signal appearing after decay of the PB and SE bands, suggesting that such signal is also present in solution, although it appears at very long delay times. Based on these results, and on new experimental evidence acquired for DBOV derivatives very recently[25], we attributed the photoinduced absorption in the near infrared to charge induce absorption due to the occurrence of ultrafast intermolecular charge-separation in the solid state, promoted by the effective supramolecular packing of such planar molecules[26]. In this scenario, the presence of such signal in solution can be linked to charge-pairs delocalized in DBOV π-aggregates, such as dimers and/or timers. Such an effect can be exploited to suppress the SE signal, especially in quantum confined systems such as nanographenes, in which ultrafast charge recombination allows the effective recovery of the emission (the on-state) in a sub-picosecond time-regime (i.e. as it has been observed in isolated polyfluorene chains[13b], see above). To achieve this in analogy with the aforementioned studies, we set-up a pump-push-probe experiment[7], in which a delayed push beam at 800 nm (i.e 1 ps of after the pump excitation) causes the re-excitation from $S_1$ to $S_n$ level promoting charges generation (see figure 6 for the experimental scheme).

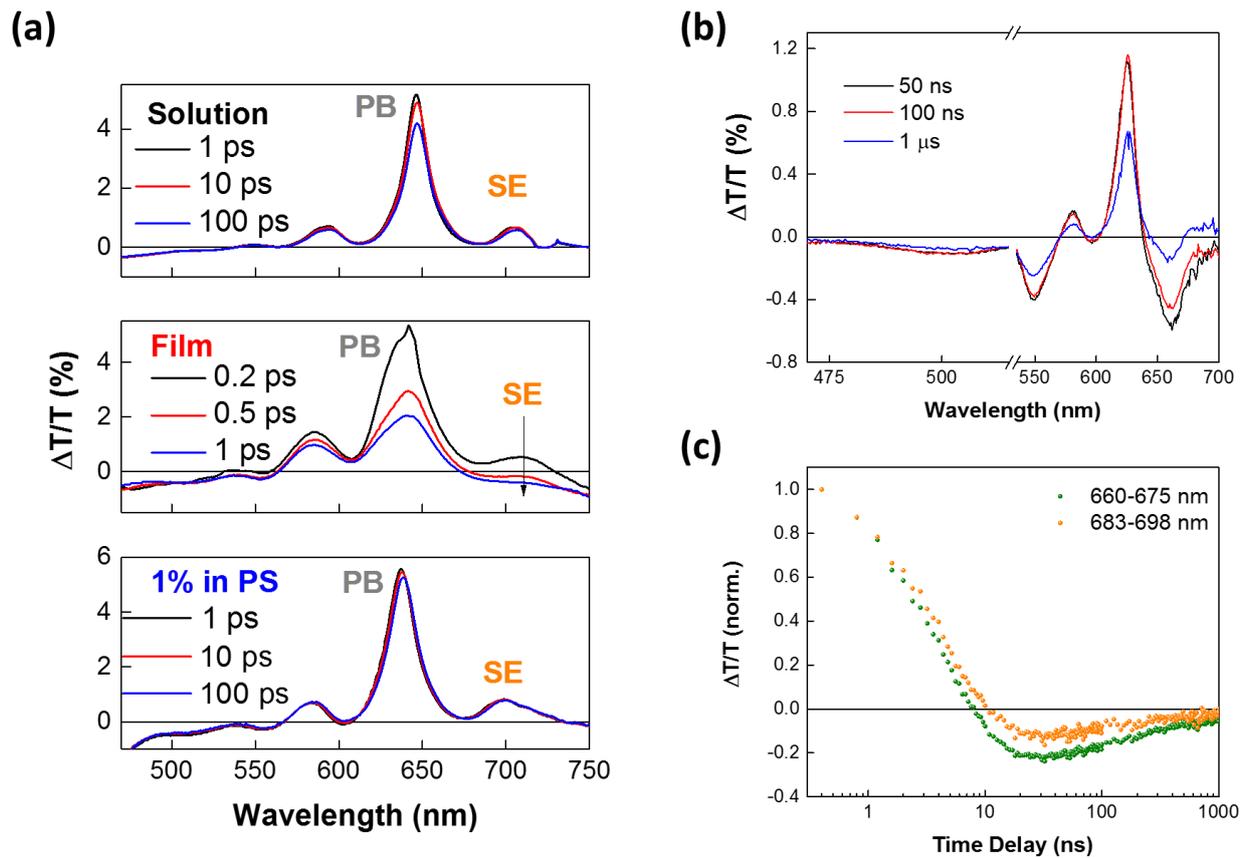

**Figure 5.** (a) Ultrafast pump-probe spectra at three pump-probe delays of DBOV toluene solution, film and DBOV:PS blend at 1 wt%. (b) Long- delay pump-probe spectra for DBOV in toluene solution showing the kicking in of a long-lived negative PA signal after depopulation in the gain region (*x*-axis in logarithmic scale). Long-delay (till 1 µs) transient dynamics at probe wavelengths corresponding to the gain region of DBOV We attribute this feature to charge pairs delocalized in DBOV aggregates. Figure 3a has been re-adapted from reference[23] under permission of John Wiley & Sons.

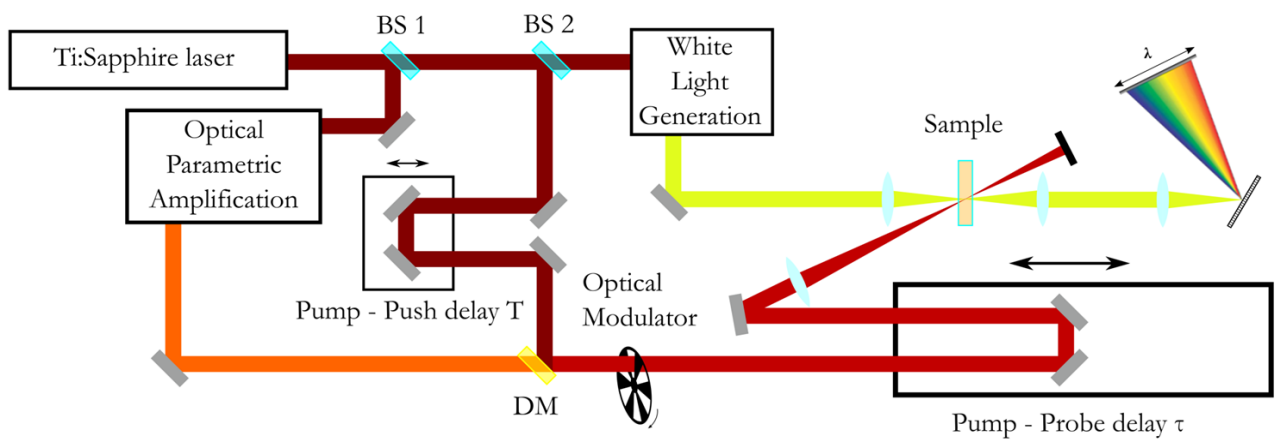

**Figure 6**. Scheme of a pump-push-probe experimental set-up. In our experiment we excited at 625 nm (pump) re-excited at 800 nm (push) and probed with a white-light continuum (probe).

To investigate the impact of the intermolecular distance on the SE switching dynamic of DBOV, we performed conventional ultrafast pump-probe and three-beams pump-push-probe (PPP) spectroscopy in solution 0.1 mg/mL and 0.01 mg/mL in toluene, and DBOV:PS blend at 1 *wt* % (figure 7). In our experiment, we pumped in resonance with the main $\pi \rightarrow \pi^*$ transition at 625 nm, pushed at 800 nm and probed with a white light continuum (430-770 nm). The differential transmission spectra for the three samples (figure 7a,c,d) show clearly a dramatic quenching of the SE upon re-excitation with the push beam for all the samples (pump-push delay at 1 ps, push excitation energy 7μJ). The effect of the push beam is in-fact two-fold: i. depletion of the $S_1$ population and ii. creation of a large photoinduced absorption band due to charges absorption. Note that the spectral shape of the push-induced negative band is different from solution (both 0.1 and 0.01 mg/mL, figure 7a,e) to the DBOV:PS blend (figure 7c), with the former peaking at 645 nm and the latter featuring a broad band shape covering almost the whole transient spectrum (500-750 nm). This might be related to different geometrical reorganization of the molecule upon push excitation when passing from solution to the rigid polymer matrix, although this has not fully investigated yet and constitutes matter for further studies. Dynamics of the SE signal at 700 nm upon push modulation at approximately 1 ps is shown in figure 7b,d,f for the three samples respectively. At this push excitation energy (7μJ), in particular, we not only switched off by 100% the SE signal, but we also achieved a gain-loss regime due to the transient dominance of the PA with the respect to SE. On the other hand, the recombination dynamic shows significant differences among the samples: whereas in the 0.1 mg/mL solution (figure 7b) the recombination process is uncomplete and occurs in ≈ 500 fs (rise time 400 fs), in the PS matrix doped with DBOV and the 0.01 mg/mL solution (figure 7d,f) the signal recovers totally within a time-frame comparable to our pulse width (150 fs). We attribute the incomplete recovery of the SE signal to charge pairs delocalization in DBOV aggregates, as a consequence of the strong tendency of such systems to form π-stacks[27], thus corroborating the long-delay pump-probe results. In this scenario, the increased PA signal outside the SE region (645 nm, blue curve in figure 2b) after push arrival it is related to the absorption of charge pairs delocalized within the aggregates. On the other hand, both in the doped polymer matrix and diluted solution we probe mostly isolated molecules, in which it is unlikely to find stabilized charge pairs. Here, we speculate that the push beam probably permits to access a state with charge-transfer character (i.e. $S_n$) whose absorption overlap with the SE region, and that can be a precursor to free charges[28] similarly to what has been demonstrated in oligofluorenes[9]. The CT state absorption spectrum resembles that of the charge pair, explaining the similar spectral competition with SE, albeit on a very short time scale associated to the intramolecular CT lifetime.

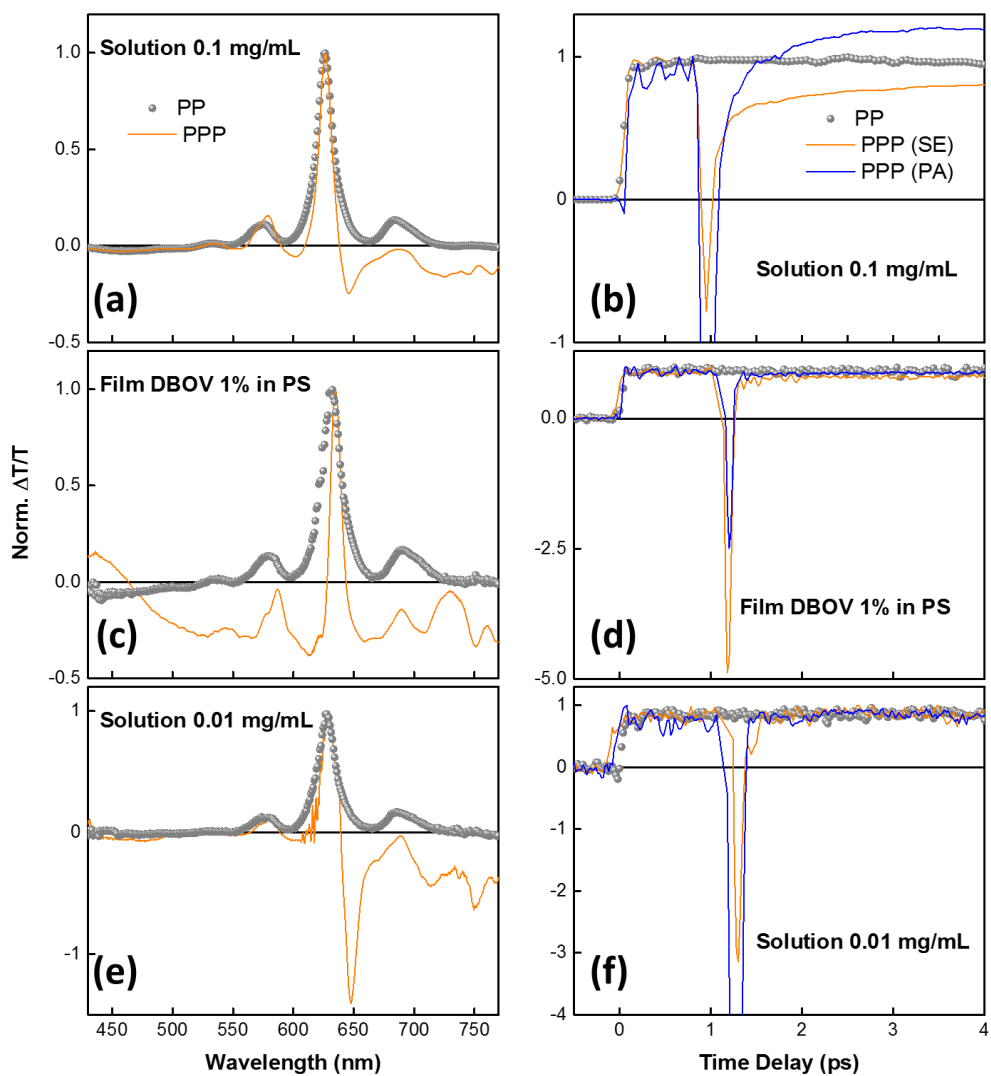

**Figure 7.** (a,c,e) Transient absorption spectra for conventional PP (grey spheres) and three-beams PPP (orange line) technique for DBOV in toluene solution at 0.1 mg/mL and 0.01 mg/mL and DBOV:PS blend at 1 *wt* % . We pumped at 625 nm, pushed at 800 nm (Ti:Sa laser fundamental) and probed with a white light continuum. The PP and PPP spectra are taken at a pump-probe delay and pump-

push delay of 1 ps. (b,d,f) Dynamic of the SE signal (orange line) and PA (645 nm, blue line) after re-excitation with the push beam (excitation energy = 7 µJ) for the three samples.

In figure 8, we show the SE dynamics of DBOV-doped polystyrene films as a function of push energy (figure 8a) and pump-push delay time (figure 8b). These films are of practical interest for possible future applications of nanographene systems in plastic optical fibers (POFs) and logics. As it has been anticipated in the previous paragraph, we observe a gain-loss regime at the highest push energies that amplifies the switching magnitude, whereas by pushing with an energy of 1 µJ we could not access that regime but still achieved a relatively strong SE quenching (86%). In addition, interestingly, we note a similar switching off/on behavior up to 10 ps pump-push delay (figure 8b) that is due to the stability of the SE signal in such a molecule (SE lifetime ≈ 400 ps[25]).

Note that although the switching process is particularly efficient and rapid for DBOV thanks to excitation confinement, this is still a resonant process that deposits a considerable amount of energy onto the material (see above). Therefore, despite our system can in principle operates into the terahertz regime, stability and energy dissipation issues have to be taken into account seriously when exciting at those frequencies. In our case at 1 THz, assuming one use the minimum switching energy for achieving 10% SE modulation (0.1 µJ), the power to be dissipated is ≈ 100 KW. Our data on ASE photostability under 2 kHz laser irradiation indicate that DBOV can easily dissipate more than 1 KW in air. Thus, we can achieve 10 GHz at these conditions. Interestingly, this further suggests that encapsulation and refinement in sample preparation/manipulation might open the way for possible operation in the THz regime.

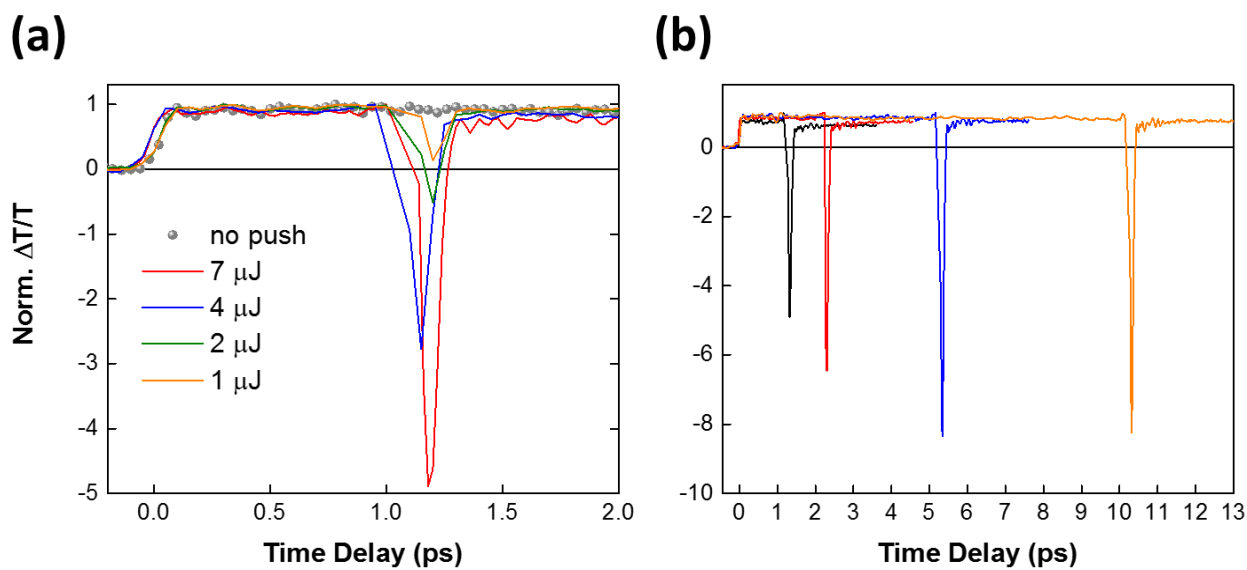

**Figure 8**. (a) Dynamic of the SE signal at various push energies and (b) at a push energy of 7 μJ at various pump-push delays for the DBOV:PS blend at 1 *wt* %.

## 5. Conclusions

In this feature article, we highlighted the contribution of PPP to investigate the complex photodynamics of conjugated molecules and to achieve efficient ultrafast all-optical switching. As a spectroscopic tool, PPP has allowed to gain insights into the charge generation process in conjugated molecules, identifying a strong spectral overlap between optical gain and charge absorption. Such an effect that can be detrimental for most photonic applications (i.e. lasers), can be exploited for modulating the gain signal. In this case, the push beam permits to access charged states whose absorption quenches stimulated emission and gain. If the conjugated system is adequately isolated, intramolecular geminate recombination of charges allows ultrafast (sub-picosecond) and effective on/off resonant switching. However, the resonant process requires an efficient energy dissipation, thus posing critical stability issues in the view to employ conjugated molecules in real devices. In these regards, graphene molecules can represent a valid alternative to conjugated poly(oligo)mers, owing to their great photostability. Interestingly, we showed that a newly synthesized nanographene molecule, namely DBOV, can sustain efficient and ultrafast resonant gain switching. The relatively high photostability displayed by DBOV and, more in general, by nanographenes can be a key asset for future applications of these novel functional materials in POFs and photonics

## 6. Experimental section

*Samples preparation.* The synthesis of DBOV was carried out through the method which we reported previously[23]. For the pump-probe and pump-push-probe measurements in solution, the molecule was dissolved in toluene with a concentration of 0.1 and 0.01 mg mL$^{-1}$. For DBOV: polystyrene solid blends (1% weight ratio), we dissolved the proper amount of material in a 40 mg/mL polystyrene solution in toluene (PS, Aldrich, Mw = 200,000). Then, the blends were spin-cast onto a glass substrate with a spin speed of 1000 rotations per minute yielding a thickness of ≈ 400 nm, as measured by profilometer.

*Pump-probe and pump-push-probe measurements.* We employed an amplified Ti:sapphire laser with 2 mJ output energy, 1 kHz repetition rate, pulse width of ≈ 150 fs and a central energy of 1.59 eV (800 nm). We used a pump wavelength of 610 nm, which is resonant with the main π → π* transition. Such pump pulses were generated by using a visible optical parameter amplifier (OPA). As push pulse we used the Ti:sapphire laser fundamental (800 nm). As a probe pulse, we used a broadband white light super-continuum generated in a sapphire plate from 450 nm to 780 nm. The pump-probe and pump-push delay were set by means of two mechanical delay stages. The pump-push-probe set-up is depicted in figure 6.

*Long-delay pump probe.* For the long-delay pump-probe spectroscopy, the pump light is obtained from a Q-switched Nd:YVO$_4$ laser (fundamental wavelength 1064 nm). This laser is electronically triggered and synchronized to the Ti:sapphire laser (which continues to provide the probe light pulses) via an electronic delay. The Q-switched pulses have a width of ~700 ps FWHM, and the system has a combined time resolution and jitter of approximately 200 ps. We pumped at 532 nm by employing a frequency doubling crystal.

## Acknowledgements

This is the pre-peer reviewed version of the following article: *Pump–Push–Probe for Ultrafast All-Optical Switching: The Case of a Nanographene Molecule*, which has been published in final form at 10.1002/adfm.201805249. This article may be used for non-commercial purposes in accordance with Wiley Terms and Conditions for Use of Self-Archived Versions.

We thank the financial support from the EU Horizon 2020 Research and Innovation Programme under Grant Agreement N. 643238 (SYNCHRONICS) and the Max Planck Society. We acknowledge fruitful discussion with Prof. C. Gadermaier.